\documentclass[prd,aps]{revtex4} 
\usepackage{graphicx}

\begin{document}
\title{Loss of entanglement in quantum mechanics due to the
use of realistic measuring rods}

\author{Rodolfo Gambini}
\affiliation{Instituto de F\'{\i}sica, Facultad de Ciencias, 
Universidad
de la Rep\'ublica, Igu\'a 4225, CP 11400 Montevideo, Uruguay}

\author{Rafael A. Porto\footnote{
Present address: Department of Physics, University of California,
Santa Barbara, CA 93106}}
\affiliation{Department of Physics, Carnegie Mellon University,
Pittsburgh, PA 15213}

\author{Jorge Pullin}
\affiliation{Department of Physics and Astronomy, 
Louisiana State University, Baton Rouge,
LA 70803-4001}

\date{September 19th. 2007}

\begin{abstract}
We show that the use of real measuring rods in quantum mechanics
places a fundamental gravitational limit to the level of entanglement
that one can ultimately achieve in quantum systems.  The result can be
seen as a direct consequence of the fundamental gravitational
limitations in the measurements of length and time in realistic
physical systems. The effect  may have implications for  
long distance teleportation and the measurement problem in quantum
mechanics.
\end{abstract}

\maketitle

Quantum field theory is ordinarily formulated in terms of
wavefunctions $\Psi(\vec{x},t)$ where $\vec{x}$ and $t$ are assumed to be
classical parameters that can be measured with arbitrary accuracy. In
reality, $\vec{x}$ and $t$ will be determined through the measurement of
physical quantities that correspond to operators. Therefore they will
have uncertainties in their measurement. In a series of recent papers
\cite{deco} we have explored this issue, in particular using
fundamental limits on the level of accuracy of measurement of
distances and times. Surprisingly, the fundamental limits are
gravitational in origin, since to measure things with very high
accuracy one requires large amounts of energy and therefore gravity
becomes relevant. As an example of the type of effects encountered, in
quantum mechanics formulated with real clocks, evolution is not
unitary, since as time evolves the real clocks will fail to mirror
exactly the evolution of the parameter $t$ that appears in the
Schr\"odinger equation \cite{Gambini:2004de}. In quantum field theory
in addition to this effect one has others since both time and space
must be measured using real devices \cite{decosp}.  In this paper we
would like to discuss another effect that arises due to the use of
real clocks and rods in quantum theory: that one cannot achieve the
maximum level of entanglement in a quantum system. Entanglement is a
fundamental effect of quantum mechanics through which a composite
system manifests non-local correlations in space. A maximally
entangled state is one in which one obtains the maximum violation possible
for a system of Bell inequalities. The latter can be considered as
the limit of correlations achievable by systems that behave
classically and locally. We will show that the level of entanglement
that one can achieve in a system is less than the usual quantum
mechanical limit due to limitations in our ability to measure
distances, and that the effect increases with the distance between the
components of the system, and also with respect to the distance to the
observer.

In order to illustrate this issue we consider a non relativistic
electron field.  We will assume we have a detector that can measure
the spin of the electron field within a certain region $v_{\vec{x}}$.
The detector is placed at some coordinate position $\vec{x}$ and we
assume the region $v_{\vec{x}}$ is centered at $\vec{x}$.  We do not
have direct access to the value of $\vec{x}$ but we assume we have a
system of real measuring rods that assign values represented by a
quantum field $\vec{X}(\vec{x})$.
This assigns a value for the position of certain reference point of
the detector as a function of its fiducial unobservable coordinates
$\vec{x}$.  We then set up an observable that measures the spin,
$\hat{\sigma}^z(v_{\vec{x}})$, given by
\begin{equation}
\hat{\sigma}^z (v_{\vec{x}}) \equiv \frac{1}{2} 
\int_{v_{\vec{x}}} du^1du^2du^3 (\hat{\Psi}^\dagger_a(\vec{u}) \sigma^z_{ab}
\hat{\Psi}_b(\vec{u}))
\end{equation}
with $\sigma_{ab}^z$ the Pauli matrix and $\hat{\Psi}_a(u_i)$ a field
operator for a field theory of a  Fermionic non-relativistic particle.

One wishes to assign a probability to the following properties that
arise from the measurement of the spin components by the detector: 
a) the z-component of the spin 
of an electron in the region accessible to the detector
and b) the physical position of the detector.



We denote the spin projectors associated with the detection of the
electron by $\hat{P}^z_{\vec{x}}(\epsilon)$, with $\epsilon=\pm 1$ the
electron spin along the $z$ direction. We also represent as
$\hat{P}^{\vec{x}_0}_{\vec{X}_0}$ the projectors associated to the
measurement of the physical position of the center of the detector,
$\vec{x}_0$ within a region $\Delta V_{\vec{X}_0}$ centered at
$\vec{X}_0$, and we consider a continuous spectrum for $\vec{X}$.
We also assume that the electron and measurement apparatus are
independent systems and therefore their joint density matrix is a
tensor product,
\begin{equation}
 \rho_{\rm total}= 
\rho_{A}\otimes \rho_{S},
\end{equation}
In the above expression
$\rho_A$ is the density matrix associated with the measuring
apparatus and $\rho_{S}$ is the density matrix associated with the
electron. This assumption is made in order to simplify calculations;
a more realistic treatment taking into account the interaction is possible,
but it would just add another source of noise that will contribute
further to the effects we discus; some discussion of this point is
in reference \cite{spekkens}). 

We would like now to ask the question ``what is the probability that
the spin takes the value $\epsilon_0$ given that the detector is
located at certain physical point $\vec{X}_0$''. Such question is
embodied in the conditional probability,
\begin{equation}\label{condprob}
{\cal P}(\epsilon_0|\vec{X} \in {\Delta V_{\vec{X}_0}})=\lim_{L\to\infty} 
{{{\int_{L}^{L} du^1\int_{L}^{L} du^2\int_{L}^{L} du^3\, {\rm Tr}
(\hat{P}^z_{\vec{u}}(\epsilon_0) \hat{P}^{\vec{u}}_{\vec{X}_0}\rho 
\hat{P}^{\vec{u}}_{\vec{X}_0})}} 
\over 
{\int_{-L}^{L} du^1\int_{-L}^{L} du^2\int_{-L}^{L} du^3\,{\rm Tr}
\left(\hat{P}^{\vec{u}}_{\vec{X}_0}\rho \right)}}
\end{equation}
where we have used the properties of the projectors to rearrange the
expression and the integrals
over $\vec{u}$ in the right hand side are taken over the whole space.  The
reason for the integrals is that we do not know for what value of the
fiducial coordinates $\vec{u}$ the rods will take the values $\vec{X}_0$.  
We now
use the independence of the system and measurement apparatus to write,
\begin{equation} \label{condprob2}
{\cal P}(\epsilon_0|\vec{X} \in {\Delta V_{\vec{X}_0}})=
\lim_{L\to\infty}
{{\int_{-L}^{L } du^1\int_{-L}^{L } du^2
\int_{-L}^{L } du^3\,{\rm Tr}(\hat{P}^z_{\vec{u}}(\epsilon_0) 
\rho_{S}){\rm Tr} (\hat{P}^{\vec{u}}_{X_0} 
\rho_{A})} 
\over 
{\int_{-L}^{L} du^1\int_{-L}^{L} du^2\int_{-L}^{L} du^3\,
{\rm Tr}(\hat{P}^{\vec{u}}_{\vec{X}_0} 
\rho_{A}) }}.
\end{equation}

The above expression 
may be written in terms of the probability of having measured
$\vec{X}_0$ for a given value of $\vec{u}$, defined by
\begin{equation} 
{\cal P}_{\vec{u}}({\vec{X}_0})={{{\rm Tr} (\hat{P}^{\vec{u}}_{\vec{X}_0} 
\rho_{A})} 
\over 
{\int_{-L}^{L} du^1\int_{-L}^{L} du^2\int_{-L}^{L} du^3\,
{\rm Tr}(\hat{P}^{\vec{u}}_{\vec{X}_0} 
\rho_{A})}}.
\end{equation}
that satisfies $\int{du^1du^2du^3 {\cal P}_{\vec{u}}({X_0}})=1$. 
The resulting expression for the conditional probability is,
\begin{equation} 
{\cal P}(\epsilon_0|\vec{X} \in \Delta V_{\vec{X}_0})=
\lim_{L\to\infty}
\int_{-L}^{L } du^1\int_{-L}^{L } du^2\int_{-L}^{L } du^3
\,{\rm Tr}(\hat{P}^z_{\vec{u}}(\epsilon_0) \rho_{S}) 
{\cal P}_{\vec{u}}({\vec{X}_0}).
\end{equation}

Due to the limitations in the measurements of lengths derived from the
fact that accuracy in measurement requires expending energy, which in
general relativity means distorting space-time, one cannot choose the
measurement apparatus such that ${\cal P}$ takes the form of a Dirac
delta. Some minimum width in the distribution is inevitable.  This
subject has a long history going back to Salecker and Wigner
\cite{SaWi}, continuing with the work of Ng and van Dam
\cite{ngvandam} and Lloyd and Ng \cite{lloydng}. 
Measurements of distances taking into account that energy must be spent
in performing a measurement, are gravitationally limited. If one
wishes to arbitrarily increase the accuracy of a measurement of
distance eventually one hits a limit in that the energy densities
involved would create a black hole.  Although this provides a fundamental
limit to the accuracy of measurement, it is obvious that in practice one will
have considerably larger errors in measurement with more mundane origins.
To model this one can consider
that ${\cal P}$ is a Gaussian whose spread grows with the distance $|\vec{X}|$
between the origin of the measuring rods and the detector,
\begin{equation}
{\cal P}_{\vec{u}}(\vec{X})
=\frac{1}{(\pi D(\vec{X}))^{3/2}}\exp{-(\vec{X}-\vec{u})^2 \over D(\vec{X})}
\end{equation}
with $D(\vec{X})={\ell_P}^{4/3}|\vec{X}|^{2/3}$ with $\ell_P$ given by
Planck's length, as has been
extensively discussed by \cite{ngvandam} 
and in our previous papers \cite{deco}. (A Lorentz-covariant
extension of these expressions, where essentially what happens is that
$|\vec{X}|$ is replaced by the proper interval, has been discussed in 
\cite{decosp}).  As the
dispersion grows with the distance to the reference point chosen from
which to measure distances, the conditional probabilities are not
invariant under translations. In fact if we consider a translation of
the origin of the rods by a vector $\vec a$, then ${\cal
P}_{\vec{u}}(\vec{X}_0)\neq{\cal
P}_{\vec{u}+\vec{a}}(\vec{X}_0+\vec{a})$ and therefore
\begin{equation} 
{\cal P}(\epsilon_0,\vec{X} \in {\Delta V_{\vec{X}_0}}) 
\neq {\cal P}(\epsilon_0,\vec{X} \in {\Delta V_{\vec{X}_0+\vec{a}}})
\end{equation}

This fact is at the core of the loss of entanglement we will discuss
in this paper.  This is a consequence from the fact that the
uncertainty in measurements of space intervals is a function of
distance and therefore translation invariance is broken by the
presence of an origin from which the measurements are obtained by
means of a physical device, which we call a ``real rod''.  In order to
study the effects of the fundamental uncertainty in position on
entangled systems we compare Bell's inequalities violations for a
given position of the detectors as measured from different origins of
the measuring rods coordinates.

Let us consider a two particle entangled system at a given time $t_0$ 
represented by the state,
\begin{equation}
|\Psi_{12}>|_{t_0}=
\frac{1}{\sqrt{(2V(v_1)V(v_2)}}(|v_1,+;v_2,-\rangle 
+ |v_1,-;v_2,+\rangle)
\end{equation}
where the kets $|v_1,+;v_2,-\rangle$ represent states of two
electrons, one with spin $+$ the other with spin $-$, the first one
localized within the region $v_1$ (centered at $\vec{x}_1$), the
second localized within the region $v_2$ and where $V(v_a)$ are the
volumes of the regions $v_a$. 

In order to measure the entanglement, we recall that a sufficient
condition for the latter is the violation of Bell's inequalities. This
phenomenon also exists in quantum field theory (see \cite{tepe} for a
discussion).  We will show later on that a state that violates the
inequalities maximally when measured locally, will violate them less
strongly when the observer gets farther and farther away.

Schematically, the Bell inequalities work like this. One measures two
quantities $Q$ and $R$ in a region centered at $\vec{x}_0$ and two
other, $S$ and $T$, in a region centered in $\vec{y}_0$. In our case
we choose these quantities to be the spins in the region of the
detector, as we defined above, with eigenvalues $\pm 1$ or $0$ (since
we are in a quantum field theory, one has to allow for a no particle
state).  For definitiveness we will consider time-like localized
measurements. That is, the detectors are devised such that they
operate within a time window around $t_0$. 

The Bell inequalities are obtained by assuming that the measured
quantities depend on classical (hidden) variables and that:

a) the values of $Q,R,S,T$ exist independently of the observation (realism),

b) Alice does not disturb the results of
the measurements of Bob when making her measurements.
 
From these hypotheses one would get that,
\begin{equation}
<\hat{Q}(\vec{x}_0)\hat{S}(\vec{y}_0)+\hat{R}(\vec{x}_0)\hat{S}(\vec{y}_0)
+\hat{R}(\vec{x}_0)\hat{T}(\vec{y}_0)-\hat{Q}(\vec{x}_0)\hat{T}(\vec{y}_0)
>\leq 2.
\end{equation}
Quantum mechanics, however,  
violates this inequality and predicts values that reach
$2\sqrt{2}$. 

We choose the operators as,
\begin{eqnarray}
\hat{Q}(\vec{x}_0)&=& 
\hat{\sigma}^z(v_{{\vec{x}_0}}) = \int_{v_{{\vec{x}_0}}} du^1du^2du^3 
\hat{\Psi}^\dagger_a(\vec{u})
\sigma^z_{ab} \hat{\Psi}_b(\vec{u})\\
\hat{R}(\vec{x}_0)&=& \hat{\sigma}^x(v_{{\vec{x}}_0})\\
\hat{S}(\vec{y}_0)&=&\frac{-\hat{\sigma}^z(v_{{\vec{y}}_0})+
\hat{\sigma}^x(v_{{\vec{y}_0}})}
{\sqrt{2}}\\
\hat{T}(\vec{y}_0)&=&\frac{\hat{\sigma}^z(v_{{\vec{y}}_0})
+\hat{\sigma}^x(v_{{\vec{y}}_0})}
{\sqrt{2}}.
\end{eqnarray}
One needs to assume that the regions $v_{\vec{x}}$ and $v_{\vec{y}}$
are spatially separated \cite{tepe} when the detector at $\vec{x}$ is
measuring the particle in $\vec{x}_1$ and the detector in $\vec{y}$ is
measuring the particle in $\vec{x}_2$. If one assumes the regions are
spherical, then their diameters should be smaller than
$|\vec{x}_1-\vec{x}_2|/2$. To simplify further calculations we will
assume that the diameters are considerably smaller than the separation,
as measured by local observers.

We wish to compute expectation values of products of these operators
to construct Bell's inequality. We will consider now a more general
situation with operators localized at points $\vec{X}$, $\vec{Y}$ by
operational measurements of the position from an origin which is not
necessarily close to the experimental arrangement.  The
expectation value of the products of these operators will be of the
form,
\begin{equation}
\langle \Psi_{12}|\sigma^z({\vec{X}}) \sigma^z({\vec{Y}}) |\Psi_{12}\rangle
 =\int d^3x d^3y\langle 
\hat{\sigma}^z({v_{\vec{x}}})
\hat{\sigma}^z({v_{\vec{y}}}) \rangle{\cal P}_{\vec{x}}(\vec{X})
{\cal P}_{\vec{y}}(\vec{Y}).
\end{equation}

We start by computing 
\begin{equation}\label{17}
  \hat{\sigma}^z({v_{\vec{y}}}) |\Psi_{12}\rangle = 
\frac{1}{\sqrt{2 V(v_1) V(v_2)}}
\left[ 
  |v_{\vec{y}} \cap v_1, +;v_2,-\rangle 
-|v_{\vec{y}} \cap v_1,-;v_2,+\rangle
- |v_1,+;v_{\vec{y}}\cap v_2,-\rangle
+|v_1,-;v_{\vec{y}}\cap v_2,+\rangle\right].
\end{equation}

To simplify things, we will assume that the volumes are equal and
only depend on the position of the center. One then has,
\begin{eqnarray}
  \hat{\sigma}^z({v_{\vec{x}}})   
\hat{\sigma}^z({v_{\vec{y}}}) |\Psi_{12}\rangle &=&
\frac{1}{\sqrt{2 V(v_1) V(v_2)}}\left[
|v_{\vec{x}}\cap v_{\vec{y}}\cap v_1,+;v_2,-\rangle-|v_{\vec{y}}\cap v_1,+;v_{\vec{x}}\cap v_2,-\rangle\right.\\
&&+|v_{\vec{x}}\cap v_{\vec{y}}\cap v_1,-;v_2,+\rangle-|v_{\vec{y}}\cap v_1,-;v_{\vec{x}}\cap v_2,+\rangle\nonumber\\
&&-|v_{\vec{x}}\cap v_1,+; v_{\vec{y}} \cap v_2,-\rangle+|v_1,+; v_{\vec{x}} \cap v_{\vec{y}}\cap v_2,-\rangle\nonumber\\
&&\left.-|v_{\vec{x}}\cap v_1,-; v_{\vec{y}} \cap v_2,+\rangle+|v_1,-; v_{\vec{x}} \cap v_{\vec{y}}\cap v_2,+\rangle\right].\nonumber
\end{eqnarray}

Notice that if one takes $v_{\vec{x}}=v_1$ and $v_{\vec{y}}=v_2$ and
${\cal P} =\delta$ (local measurement) then, using equation (\ref{17})
one has that $\vec{X}=\vec{x}_1$ and $\vec{Y}=\vec{y}_1$ and one is
left only with the fifth and seventh term since $v_1$ and $v_2$ are
disjoint and,
\begin{equation}
\langle \Psi_{12}|\hat{\sigma}^z(\vec{X})
\hat{\sigma}^z(\vec{Y}) |\Psi_{12}\rangle= -1,
\end{equation}
which is the usual result in quantum mechanics.

In general, for arbitrary $v_{\vec{x}}$ and $v_{\vec{y}}$, 
\begin{eqnarray}
  &&\langle \Psi_{12}|\hat{\sigma}^z(\vec{X})
\hat{\sigma}^z(\vec{Y}) |\Psi_{12}\rangle=\int d^3x \int d^3 y 
\frac{{\cal P}_{\vec{x}}(\vec{X})
{\cal P}_{\vec{y}}(\vec{Y})}{2 V(v_1) V(v_2)}
\nonumber\\
&&\times\left[2 V(v_2) V(v_{\vec{x}}\cap v_{\vec{y}} \cap v_1)
-2 V(v_{\vec{y}}\cap v_1) V(v_{\vec{x}}\cap v_2)- 
2 V(v_{\vec{x}}\cap v_1) V(v_{\vec{y}}\cap v_2)
+2V(v_1) V(v_{\vec{x}}\cap v_{\vec{y}} \cap v_2)\right]\\
&=&\int d^3x \int d^3 y 
{{\cal P}_{\vec{x}}(\vec{X})
{\cal P}_{\vec{y}}(\vec{Y})}\left[
\frac{V(v_{\vec{x}}\cap v_{\vec{y}} \cap v_1)}{V(v_1)}
+\frac{V(v_{\vec{x}}\cap v_{\vec{y}} \cap v_2)}{V(v_2)}
-\frac{V(v_{\vec{x}}\cap v_1) V(v_{\vec{y}} \cap v_2)}{V(v_1)V(v_2)}
-\frac{V(v_{\vec{x}}\cap v_2) V(v_{\vec{y}} \cap v_1)}{V(v_1)V(v_2)}\right].
\nonumber
\end{eqnarray}

It is also the case that
$\langle \Psi_{12} |
\hat{\sigma}^z(\vec{X}) \hat{\sigma}^x(\vec{Y})
|\Psi_{12}\rangle=0$ since
after the action of these operators the spins are either
 are $++$ or $--$ and therefore orthogonal to the
initial ones in $|\Psi_{12}\rangle$. For the $x$ direction we have,
\begin{eqnarray}
\langle\hat{\sigma}^x(\vec{X})
\hat{\sigma}^x(\vec{Y})\rangle&=&
\int d^3x \int d^3 y 
{{\cal P}_{\vec{x}}(\vec{X})
{\cal P}_{\vec{y}}(\vec{Y})}\left[
\frac{V(v_{\vec{x}}\cap v_{\vec{y}} \cap v_1)}{V(v_1)}+
\frac{V(v_{\vec{x}}\cap v_1)V(v_{\vec{y}}\cap v_2)}{V(v_1)V(v_2)}\right.
\nonumber\\
&&\left.+
\frac{V(v_{\vec{x}}\cap v_2)V(v_{\vec{y}}\cap v_1)}{V(v_1)V(v_2)}+
\frac{V(v_{\vec{x}}\cap v_{\vec{y}} \cap v_2)}{V(v_2)}\right].
\end{eqnarray}

To set up Bell's inequalities we need the expectation values,
\begin{eqnarray}
\langle\hat{Q}(\vec{X})\hat{S}(\vec{Y})\rangle&=&  
-\frac{1}{\sqrt{2}}\langle\hat{\sigma}^z(\vec{X})
\hat{\sigma}^z(\vec{Y})\rangle\\
\langle \hat{R}(\vec{X})\hat{S}(\vec{Y})\rangle&=&   
\frac{1}{\sqrt{2}} \langle\hat{\sigma}^x(\vec{X})
\hat{\sigma}^x(\vec{Y})\rangle\\
\langle\hat{R}(\vec{X})\hat{T}(\vec{Y})\rangle&=&   
\frac{1}{\sqrt{2}} \langle\hat{\sigma}^x(\vec{X})
\hat{\sigma}^x(\vec{Y})\rangle\\
\langle \hat{Q}(\vec{X})\hat{T}(\vec{Y})\rangle&=&   
\frac{1}{\sqrt{2}} \langle\hat{\sigma}^z(\vec{X})
\hat{\sigma}^z(\vec{Y})\rangle,
\end{eqnarray}
so one has
\begin{equation}
\langle\hat{Q}\hat{S}+
\hat{R}\hat{S}+
\hat{R}\hat{T}-
\hat{Q}\hat{T}\rangle=
\frac{4}{\sqrt{2}}  \int d^3x \int d^3 y 
{\cal P}_{\vec{x}}(\vec{X}){\cal P}_{\vec{y}}(\vec{Y})
\left[
\frac{V(v_{\vec{x}}\cap v_1)V(v_{\vec{y}}\cap v_2)
+V(v_{\vec{x}}\cap v_2)V(v_{\vec{y}}\cap v_1)}{V(v_1)V(v_2)}\right]\label{27}
\end{equation}

Therefore for ${\cal P}=\delta$ and $v_{\vec{x}}=v_1$ and
$v_{\vec{y}}=v_2$ one has that
\begin{equation}\label{28}
\langle\hat{Q}\hat{S}+
\hat{R}\hat{S}+
\hat{R}\hat{T}-
\hat{Q}\hat{T}\rangle =2{\sqrt{2}}, 
\end{equation}
as in standard quantum mechanics, leading to a maximal violation of
Bell's inequality. This result corresponds to a local measurement
where the origin of measurement is close to the particles and the particles
are close to each other, so one does not have to worry about the
dispersion we are studying.

Let us now consider the situation where these hypotheses do not hold
anymore. That is, we will consider $x_1$ and $x_2$ to be close to each
other  but we will measure the position of the
detectors from an origin that is at a large distance $a$ from the
detectors. So we will now consider functions
$\hat{Q}(\vec{X})$ with $\vec{X}\sim {a}$, etc.  Let us take $\vec{x}_1$ and
$\vec{x}_2$ to be the position of $v_1$ and $v_2$ (measured
locally). When measuring the system from a distance $\vec{X}$
and $\vec{Y}$ such that $\sqrt{D(\vec{a})}\sim
|\vec{x}_2-\vec{x}_1|$ (with
$D(\vec{a})={\ell_P}^{4/3}|\vec{a}|^{2/3}$ as before), we will have to
include in the average of equation (\ref{27}) regions where $v_x\cap
v_1$ and $v_x\cap v_2$ (and similarly for $v_y$) are empty. This
immediately implies that the right hand side of Bell's inequality
(\ref{28}) will be smaller than $2 \sqrt{2}$. It should also be noted
that when measuring the system from a distance $\vec{a}$ such that
$\sqrt{D(\vec{a})}\sim r(v_{\vec{x}})$ or that $\sqrt{D(\vec{a})}\sim
r(v_{\vec{y}})$ where $r(v_{\vec{x}})$ is the characteristic size of the
region $v(\vec{x})$ and similarly for $\vec{y}$, the effect will also
be large. This is due to the fact that we will have to include regions
where $v_{\vec{x}}\cap v_1$ and $v_{\vec{y}}\cap v_2$ are empty. In most
practical applications this will be in fact the dominant
effect. However, this effect can always be diminished by increasing
the size of the detectors.

We could introduce operators with $v_{\vec{x}}$ and $v_{\vec{y}}$ much
lager than $v_1$ and $v_2$ and in that case the inequality would still
be violated if we only take into account the previous arguments,
although it wont achieve the value $2\sqrt{2}$.  The combined effect
of both contributions is again the elimination of the correlation
between both particles. In other words the entanglement observed by
different observers is different and the degree of entanglement
diminishes as the system is observed from further and further away. In
this sense we can say that the entanglement {\em is not} translation
invariant. It is known that the entanglement of spins is not Lorentz
invariant \cite{terno}. In this case this is an additional effect.

The effect we are considering, for a system of two entangled
particles, will always be very small for measurements done close to
the particles (that we assume are not far apart from each other),
which are the usual kinds of measurements considered for entangled
systems. Therefore, even though in practice there is loss of
entanglement due to a translation to a different observational point,
the entanglement can be always recovered by choosing a local observer.


The effect becomes more interesting when one considers more than two
entangled particles. There one can have more favorable
configurations. Consider a system of four particles consisting of two
pairs of particles close to each other within each pair, whereas the
pairs are far away from each other.  Let us call the separation of the
particles within each pair $\delta$ and the separation between pairs
$d$. We then have that $d\gg\delta$. In this case our effect will be
of increasing importance if $D(d)\sim \delta^2$.
The effect would then go as,
\begin{equation}
\exp\left[-\left(\frac{\delta}{d^{1/3} 
\ell_{\rm Planck}^{2/3}}\right)^{2}\right].
\end{equation}

So for instance, if one considered $\delta$ of the order of nuclear
distances, one would require $d$ to be of the order of millions 
of kilometers for the effect to be of the order of $10^{-6}$.
This cannot be completely discarded in the context of experiments
involving interferometry in space.

We have shown that the use of realistic measuring rods in quantum
mechanics inevitably leads to a loss of entanglement. The effect
depends on the distance from the measured system to the origin 
chosen and goes as the distance to the one third power. The magnitude of
the effect is very small for everyday experiments, but could put limitations
to teleportation over astronomical distances. 

Another interesting implication of the effect is for the measurement
problem in quantum mechanics \cite{despagnat,zurek,foundations}. In
the usual treatment of the measurement problem, one considers
decoherence due to interaction with the environment. In these effects
one assumes that simultaneous measurements of the system, the
measurement apparatus and the environment are impossible due to
practical limitations having to do with the large number of degrees of
freedom of the environment.  The existence of the loss of entanglement
we discuss would add fundamental (and not only practical) limitations
to the simultaneous measurement of system, apparatus and
environment. As such, it could bypass usual objections that suggest
that at least in principle such measurements would be possible. Our
effect provides a definite reason for the passage from the quantum to
the classical world that cannot be overridden even in principle.

This work was supported in part by grant NSF-PHY-0554793, funds of the
Hearne Institute for Theoretical Physics, FQXi, CCT-LSU and Pedeciba
(Uruguay). RAP is supported in part by DOE contracts DOE-ER-40682-143
and DEAC02-6CH03000.

\maketitle

\end{document}